WITHOUT SPECTROSCOPY AT THE BEGINNING, CATALYSIS RESEARCH PROCEEDED IN THE WRONG DIRECTION FOR MORE THAN 100 YEARS

Ralph A. Gardner-Chavis[1] John T. Reye[2], Theodore B. Selover, Jr.[3], Huixiong Zhang[4]

*1. Chemistry Department, Cleveland State University*

*2. School of Chemical and Biomolecular Engineering, Georgia Institute of Technology*

*3. Cuyahoga Community College*

*4. Diapedia, LLC, State College ,Pennsylvania*

## ABSTRACT

One hundred sixty years ago without the benefit of the appropriate instrument, researchers in catalysis came to the only possible logical conclusion at that time. Namely, that the reactants, intermediates, and products are attached to the catalyst. The steps of the standard experimental procedure became: 1. prepare the catalyst, 2. expose it to a gas for some period of time, 3. thoroughly evacuate the system to remove any vestige of physisorption, 4. study the chemisorbed species by weighing with a micro-balance or other method. After some 80 years of this procedure, when infrared spectroscopy, the instrument that could identify molecules on surfaces became available, it was so automatic to "evacuate before investigate" apparently no one examined the physisorption region of the experimental procedure until this present author in 1954. Then, the first experiments demonstrated that when the surface is a catalyst, carbon monoxide produced several surface species at frequencies deviating more than 200 $cm^{-1}$ from the gas frequency. While these surface species are in a gas-like state, they desorbed only as normal carbon monoxide gas. This was the first major discovery from this study. **It is because these intermediates of catalysis exist only on the surfaces of catalysts and are readily removed by evacuation that they were not detected prior to this study in 1954. The field of catalysis must accept and incorporate this information into its procedures to accelerate its advances.** New reactions and catalysts may be discovered serendipitously by present techniques because reactants attached to the catalyst may detach unobtrusively and proceed to products as described in the following. Gardner-Chavis, using the infrared technique pioneered by R.P. Eischens, observed that these physisorbed surface species on catalysts have frequencies intermediate to the frequencies of ions and neutral molecules. Thus these frequencies correspond to non-integral numbers, the fraction of which is unique for each catalyst. The Atomic Energy Level tables from NIST list available energy states for electrons excited out from the ground state of atoms and positive ions of almost all of the elements For example, the first state above the ground state for the chromium atom is at 7,593.16; the first state for the positive ion is at 11,961.81; the first states for the double positive ion are at 62.22, 183.16, 356.55, and 576.08 $cm^{-1}$. Excited states this close to the ground state will be occupied for some fraction of time thus creating an electric field with a fractional charge. Each positive ion that may function as a catalyst has a different set of these low-lying levels. Thus each ion presents a different fractional field to prepare the chemical reactants for the catalytic event. Electric fields with a fractional charge were the second major discovery of this study.

The.catalytic site for a reaction of the type A + B ↔ C is composed of two identical ions, one for each reactant. While these two ions must not be connected by a chemical bond they must be close enough that their fields create a region of overlap. The catalyst which is appropriate for a particular reaction has electric fields with the fractional charge that when this charge is imparted to the two reactants changes the frequency of the electric fields of the reactants at the sites of reaction so that they now match harmonically. That is, the ratio of the frequency of one to the other equals a power of two. As the reactants pass through the overlap region, because their frequencies now match, resonance occurs. It is during resonance that electrons are shared, paired and exchanged and bonds are broken and made to satisfy the Free Energy constraint. **That the catalyst is the electric fields explains the unusual observation that the catalyst is not consumed by use.**

The concept of electric fields operating in concert to control mechanical and chemical reactions is the third and most extraordinary discovery of this study Applications to and explanations of many phenomena in mechanics and biology as well as chemistry may result from this discovery. Quantitative mathematical analyses of reactions are presented for oxidation, chlorination, carbonic anhydrase homologues, and the elimination of nitric oxide from auto exhaust.

## DISCUSSION

It is difficult to introduce a completely new and different paradigm into a field, such as catalysis, that has been established for 160 years. Thus it will be necessary to present in this paper not only previously published examples of quantitative applications of the new relations to catalysis but also one or more applications to other related areas of chemistry. First some misconceptions must be cleared up.

Just for a moment, divide catalysts into to two groups, inorganic catalysts and organic catalysts. Inorganic catalysts are metals and metallic compounds. Some are found naturally, but most are compounded for industrial applications. A single piece of catalytic material may contain hundreds to millions of active catalytic sites. These catalysts may function as solids or in solution and are simply called "catalysts".

The organic catalysts are called "enzymes" and are the catalysts in living organisms. Enzymes generally are a single site composed of two identical ions. Some and possibly all enzymes have proteins associated with them. The proteins facilitate the activity of the enzyme catalysts in many ways by orienting and presenting reactants and removing products. A brief discussion concerning the proteins will be given later.

There has been ongoing confusion because some enzymes contain metals and some do not. This has lead some to call the metals in enzymes 'co-factors' and others to assume that the proteins were the catalysts. The explanation is that the catalytic event depends, not on the presence of a metal, but upon the overlap of electric fields with fractional charges. Data in the Atomic Energy Level (AEL) tables show certain ions that are not metallic contain the low-lying excited electron levels above the ground state necessary to generate electric fields with fractional charges. These are certain positive ions of carbon, nitrogen, oxygen, phosphorus, and sulfur that therefore satisfy the criterion for catalytic activity. There are other metals and possibly non-metals in various oxidation states in the AEL tables that function as catalysts that do not contain the low-lying excited electron levels. Since the AEL tables list only positive ions it is considered that these ions must be functioning as negative ions. It is interesting that several of these metals such as sodium, zinc, silver, cadmium, gold and magnesium are associated with living organisms

Everyone is familiar with steam condensing upon a mirror. If the temperature is raised above the "dew-point", the water will evaporate from the surface. Water went onto the surface and water comes off the surface. There is no reason to think that the water may have existed on the surface in any form other than as water. Early in the field of catalysis it was concluded that there were only two possible results from a gas condensing upon a surface. One reaction, known as chemisorption, is the formation of a chemical bond between the molecule from the gas phase and an atom embedded in the surface. The other reaction, known as physisorption, is demonstrated by the condensation of steam on a mirror; a reaction so gentle that the water molecule is changed almost imperceptibly. Physisorption is much too weak to account for the large changes in reactivity encountered in catalysis. Therefore the scientists were left with chemisorption as the only possible reaction of importance to catalysis. **Without instrumentation capable of examining molecules on the surface of a solid, there was no way that investigators could have known that atoms and molecules on catalysts were converted in many cases to as many as four different species that turned out to be the long-sought intermediates of catalysis.** Because the frequencies of these surface species are intermediate to those of the neutral molecule and its positive and negative ions, these surface species were named 'intermedions'. Intermedions exist only on the surfaces of catalysts. They are in a gas-like state and are immediately removed in the form of the original gas. Therefore, for 160 years, standard experimental technique carefully removed intermedions before studying the chemisorbed ones. The frequency referred to in this paper is the frequency portrayed in infrared spectroscopy which is the difference in frequency between two states of the electric field.

.

Eischens[1] pioneered in the use of infrared spectroscopy to examine chemisorbed species on solid surfaces. In 1954, the initial experiments in this study by Gardner-Chavis[2] for his doctoral thesis were to use the Eischens' technique to check the physisorption region of catalysis. The surfaces used were copper oxide or a reduced cobalt oxide supported on Cab-O-Sil, a silica gel of surface area approximately 400 square meters per gram. Carbon monoxide, CO, was added to the insitue cell in increasing amounts with accompanying spectra until the doublet of gaseous CO was observed. Observing gaseous CO indicated that at least a mono-layer of CO was on the surface. Next the experimental procedure required the pressure to be decreased by 50%. According to conventional expectations, all bands in the spectrum corresponded to chemisorbed CO except a band at 2143 $cm^{-1}$, the frequency of physisorbed CO. Upon a decrease in pressure this should be the only band to disappear. However, several bands at frequencies of 2173, 2121, 2000 and 1932 $cm^{-1}$ from copper oxide and 2179, 2160 and 2091 $cm^{-1}$ from cobalt disappeared when the pressure was decreased. These infrared absorption bands were from some form of CO because the gas removed from the surface by the decrease in pressure was only CO. These surface species were on the surface in a gas-like state. G. Herzberg, Nobel Laureate[3] in spectroscopy, had published the frequencies of the neutral CO molecule with 10 outershell electrons at 2143.27 $cm^{-1}$ and the positive ion at 2183.90 $cm^{-1}$ with 9 electrons. Gardner-Chavis expected that the seven surface species that were removed by the decrease in pressure along with the data for the neutral CO molecule and the positive ion should all fit a single equation of the form of a right hyperbola, viz.,

$$(\nu - A)(\eta - B) = k \qquad (1)$$

wherein '$\nu$' represented frequency and '$\eta$' the number of electrons, 'A and B' are asymptotes and 'k' is the curvature. In addition to the data from Herzberg, the necessary third constant chosen for the point of dissociation at an integral number of electrons was 12. **The simplicity of this type of equation belies its very great utility in spectroscopy as well as catalysis.** The equation[4] for CO is:

$$(\nu_{co} - 2269.96)(\eta_{co} - 12.1182) = 268.309\,cm^{-1} \qquad (2)$$

According to Eq. 2, the numbers of electrons corresponding to the four surface species on copper oxide are: 2173 (9.35); 2121 (10.32); 2000 (11.12); 1932 (11.32) and the three on cobalt are 2179 (9.16); 2160 (9.68); 2091 (10.62). Except for the bands at 2000 on copper oxide and 2179 on cobalt, the others present, within 0.06 of an electron, a unique fraction characteristic of each metal in the solid, namely, 0.32 for copper and 0.62 for cobalt.. Equation (2) is more accurate at lower frequencies and larger numbers of electrons therefore the fraction for cobalt was considered at this stage of the research to be 0.62.

The following equations similar to Eq. (2) for CO were derived in a previous paper[5] for the three normal modes of carbon dioxide, $CO_2$. These equations for $CO_2$ were known to be approximate. However the variation in "k" between the three normal modes which from lack of data could not be determined at this time was nevertheless expected to be much less than for the other two variables.

$$(\nu_1 co_2 - 1418.2)(\eta_1 co_2 - 18.120) = 170.18\ cm^{-1} \qquad (3)$$

$$(\nu_2 co_2 - 742.1)(\eta_2 co_2 - 18.229) = 170.18\ cm^{-1} \qquad (4)$$

$$(\nu_3 co_2 - 2431.1)(\eta_3 co_2 - 18.070) = 170.18\ cm^{-1} \qquad (5)$$

.

The literature was consulted to determine if other investigators using the Eischens' technique with CO or $CO_2$ on other solids had inadvertently reported two or more IR bands which when converted to numbers of electrons by Eqs.2, 3, 4 or 5, corresponded to different integers but the same fraction. Noting that the fraction is the same whether CO or $CO_2$ is the surface species demonstrates that the fraction is a property of the metal component of the solid that is not affected by the surface gas.

**It was not always obvious from the descriptions of others whether their spectra were obtained before evacuation or after. Thus some data from other studies may be due to chemisorbed species and therefore not expected to fit the pattern**.

Data for $CO_2$ in contact with solids containing various metals are included in the lists below because, despite integers of 16 and 17, they show the same fractions as CO. The number of asterisks corresponding to the three normal modes also distinguishes bands due to $CO_2$. All bands from CO and $CO_2$ reported by thirteen other investigators on eight different surfaces are in terms of $cm^{-1}$:

2127, 2020 and 1960; Fe and Fe oxide; Eischens and Pliskin[6]
2165 and 2099; Ag and Ag oxide; Huber et al[7]
2203, 2184 and 2134; Cr oxide; Gardner[8]
2188, 2058, 1965, 1230 and 1635***; Ni and Ni oxide; Eischens[9] Little and Amberg[10] , Pickering and Eckstrom[11], O'Neill and Yates[12], Filimonov[13]

2174 and 2198***;zinc oxide; Taylor and Amberg [14]
2288***,1379***, 658.3*,657.2*,630.9*, NaCl; Kozirovsky and Folman[15]
2146, 2074 and 2040; Ru and Ru oxide; Abhivantanaporn and Gardner[16]
1807***, 356** and 850*; gold; Huber, McIntosh and Ozin[17]

Very little analysis or interpretation can be made of these data, no more than that some bands are above, between or below the data from Herzberg of 2183.90 for the positive CO ion and 2143.27 for the neutral CO molecule and below the three normal modes of gaseous $CO_2$ at: $\nu_1$(1336.9), $\nu_2$(667.3), $\nu_3$(2349.3) $cm^{-1}$.

When the spectral bands are translated by Eq. 2 for CO and the corresponding equations for $CO_2$, into numbers of electrons, the same pattern observed for the data of CO on copper oxide and reduced cobalt oxide becomes evident, namely, two or more bands on each metal corresponding to different integers but a constant fraction.

Applying Eqs. (2), (3), (4) and (5) to these data gives
10.24, 11.04 and 11.25; Fe and Fe oxide
9.56 and 10.55; Ag and Ag oxide
8.11, 9.10 and 10.14; Cr oxide
8.84, 10.85, 11.25, 11.86 and 17.86; Ni and Ni oxide
9.32 and 17.34; Zinc oxide
16.89, 17.91, 17.896, 17.899, and 17.904; NaCl
9.94 and 10.94; Ru and Ru oxide
17.80, 17.79 and 17.82; gold

The precision within 0.04 electrons of all of these fractions is substantial support for the validity of Eq. 2, the right hyperbola relating vibrational frequency as a function of the number of electrons on CO and the similar equations for $CO_2$. This precision is also a tribute to the painstaking experimental technique of these scientists who had no idea that their data would be used as above. The metallic component of the solid is clearly identified by the fraction of each intermedion. These fractions are called 'perturbation fractions, 'PF' since these are the fractional amounts by which molecules (and atoms) are altered by the electric field of the adsorbent. At this present time a four-digit number identifies some of these fractions without ambiguity. A list of four-digit fractions of ions from experiment is presented at the end of this paper. A discussion of "outliers" is presented in the Appendix.

The appearance of the symmetrical stretching normal mode, $\nu_1$ of $CO_2$ on sodium chloride and on gold was most unexpected. Since there is no dipole moment change during a symmetric stretch in the linear $CO_2$ molecule, this band is not observed in infrared spectra of gaseous $CO_2$. The appearance of this band indicates that the molecule has become bent from $D_{\infty h}$ to $C_{2v}$ symmetry as the apparent number of electrons is increased. According to Walsh[18], the addition of electrons to $CO_2$ should cause it to bend as observed.

One of two data in the above paragraph that is particularly important is the band on chromium at 2203 $cm^{-1}$ corresponding to 8.11 electrons. This band gives support to Eq. 2 to within 0.11 electrons of the double positive CO ion. Krupenie[19] reported a value of 14.0 eV as the energy necessary to remove one electron from CO. To remove a second electron from the positive ion would require in excess of 28 eV. Greater energy would be required to remove a second negative electron from an ion made positive by the removal of the first electron. The other band of extra significance is at 1230 $cm^{-1}$ for CO on nickel oxide corresponding to 11.86 electrons in agreement with 8.84 and 10.85 for CO and 17.86 for $CO_2$. This band for CO supports Eq. 2 to within 0.14 electrons of 12, the number of electrons that correspond to dissociation of CO into negative ions. **Frequencies of 2203 and 1230 $cm^{-1}$ are far beyond the range of frequencies previously reported for any CO entity.** Normal frequencies of CO range between 2183.90 $cm^{-1}$ for $CO^+$ and 1710 $cm^{-1}$ for $(C_6H_5)_2CO$. Thus, while intermedions may be physisorbed species affected by forces that are even greater than ordinarily encountered in chemistry, in all cases when tested they are not attached to the surface but are readily removed from the surface immediately by a decrease in pressure.

Theodore Selover, Jr[20]. recommended the Atomic Energy Level[21] (AEL) tables as a possible source to find data to explain the origin of these fractions, The AEL tables list the electronic states above the ground state for atoms and positive ions We looked at the chromium atom and observed the first state above the ground state to be at 7,593.16 $cm^{-1}$. The first state above the ground state for the positive chromium ion was at 11,961.81$cm^{-1}$ The first four states above the ground state for the double positive chromium ion were at 62.22, 183.16, 356.55 and 576.08 $cm^{-1}$. It was immediately obvious that **states this close to the ground state would be populated with an electron for a fraction of time thus creating an electric field with a fractional charge. Each positive ion that can function as a catalyst has a different set of these low-lying excited electron states. Thus each such ion provides an electric field with a unique charge as a site for catalysis.**

Selover[20] suggested investigating the relationship between the fractions that are characteristic of each ion and the data in the AEL tables, It was necessary to express the low-lying levels of an ion by a single number when there is more than one level. The Boltzmann function provides the sum of energy levels each weighted in accordance with its contribution to the sum

$$U = \sum_{\substack{n \neq 0 \\ n=1}}^{n=\infty} (2J_n + 1)\exp\left(\frac{-hc\overline{\nu}_n}{kT}\right) \qquad (6)$$

In Eq. (6); *h* is Planck's constant; *c*, the speed of light; *k*, Boltzmann's constant; *T*, 298.15K, and $(2J_n+1)$ is the multiplicity of each level. Since it is proposed that it is the occupation of energy levels above the ground state that creates the electric field, the ground state is not included in the sum. Thus the sum, U, is a truncated electronic partition function. The factor hc/kT was used as $4.82556\text{x}10^{-3}$.
At this point in the research[20] there were four ions that have low-lying levels in the AEL tables that have produced surface species readily removed by evacuation with IR frequencies corresponding according to Eq. (2) to fractions of electrons. These ions are III Cr, II Fe, II Co and III Ni.
The Roman numerals preceding the atomic symbol are physicist's ionization states.
The graph of these data is presented in the following Figure-1.

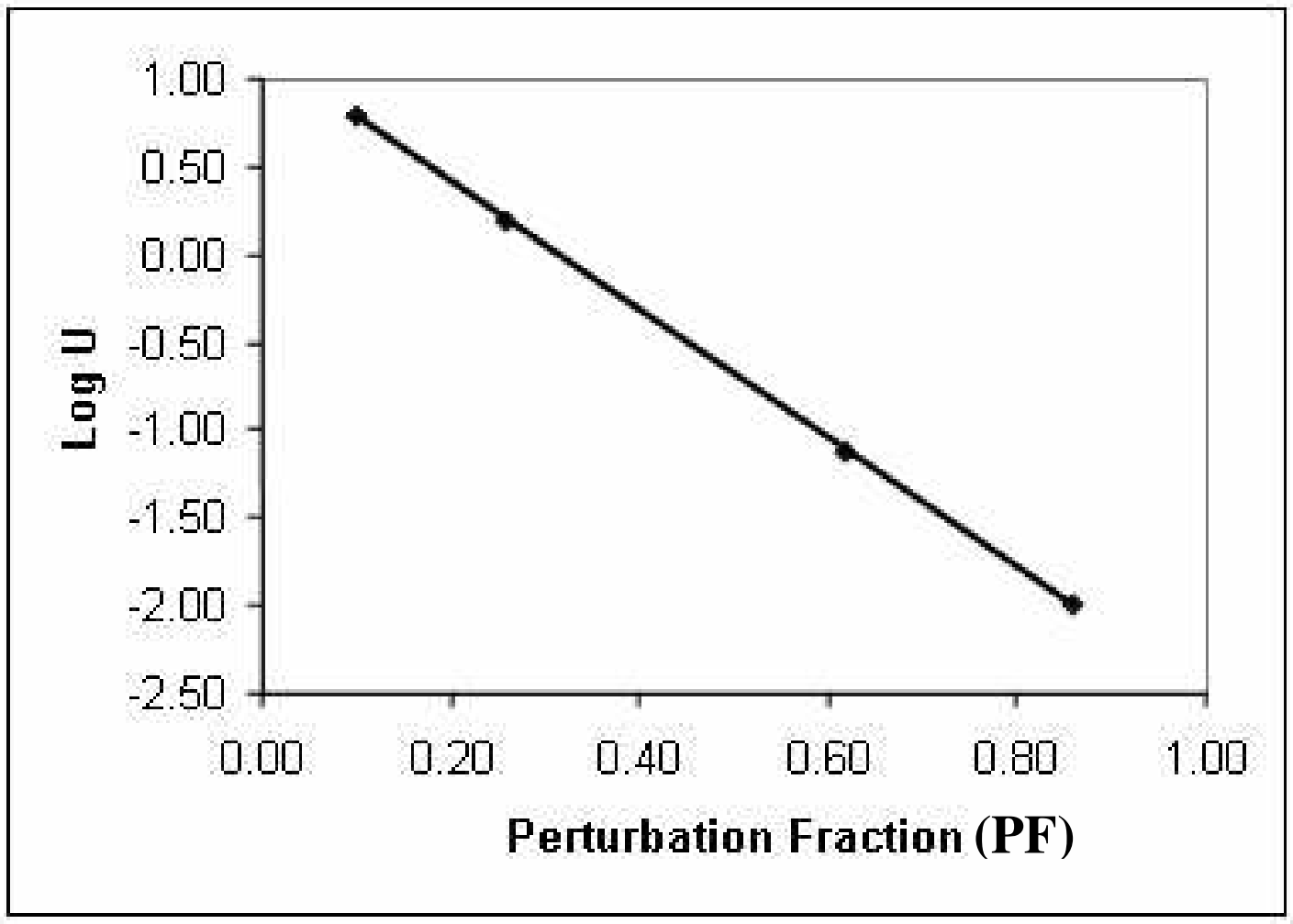

| PF | Metal Ion | U | Log U |
|---|---|---|---|
| 0.10 | III Cr | 6.099 | 0.785 |
| 0.26 | II Fe | 1.570 | 0.196 |
| 0.62 | II Co | 0.074 | -1.131 |
| 0.86 | III Ni | 0.00994 | -2.003 |

**Figure 1 Data from Atomic Energy Level tables, 'U' compared with fractions, 'PF' assigned to catalytic metals.**

The direct linear relation between log U and the PF demonstrates that the electric field of the catalyst produces the fractions that change the frequency of the electric fields of the reactants, thus leading to catalysis. In a new appropriate definition of a catalyst, the presence of an electric field with a fractional charge should be the criterion.

## EXPLANATION OF ANOMALOUS FREQUENCIES OF HYDROGEN PEROXIDE

When atoms and groups of atoms combine to create larger molecules, electrons are shared with the result that each group may have fractional possession. There is a problem in the assignment of the frequencies of the stretching modes in hydrogen peroxide[22]. In water where the two hydrogen atoms are attached to the same oxygen, the symmetrical and unsymmetrical modes are assigned at 3656 and 3755 $cm^{-1}$ with a difference is 99 $cm^{-1}$. The frequencies assigned for these modes for hydrogen peroxide are at 3417 and 3540 $cm^{-1}$, a difference of 123 $cm^{-1}$ with the hydrogen atoms attached to different oxygen atoms. It would be expected that the difference between the symmetrical and unsymmetrical modes for hydrogen peroxide would be significantly less or even zero.

G.Herzberg[3] presented data accurate to six significant figures for the frequencies of the triplet sigma neutral oxygen molecule and its positive ion. The frequency for the neutral molecule with 12 electrons is 1556.38 $cm^{-1}$ and 1843.34 $cm^{-1}$ for the positive ion with 11 electrons. Taking the point of dissociation by the addition of electrons to anti-bonding orbitals as the -2 ion with 14 electrons, the following equation for oxygen intermedions and the equation for hydroxyl intermedions from a previous publication[23]are:

$$(\nu_{OO} - 2920.16)(\eta_{OO} - 15.7525) = 5117.58\ cm^{-1} \qquad (7)$$

and

$$(\nu_{OH} - 6090.25)(\eta_{OH} - 0.16649) = -17256.8\ cm^{-1} \qquad (8)$$

The addition of two electrons to the fourteen of hydrogen peroxide, making a total of sixteen, causes the molecule to break into two OH molecules of eight electrons each. The frequency of the O-O bond of hydrogen peroxide with sixteen electrons is equal to the frequency of the O-O bond of the oxygen molecule with fourteen electrons, which is zero. The frequency of the OH bond of hydrogen peroxide with sixteen electrons is the same as the OH molecule with eight electrons. Removing six electrons from hydrogen peroxide leaves only the eight electrons in bonding orbitals as the hydrogen atoms fall off as protons. The O-O frequency of hydrogen peroxide with eight electrons is the same as an oxygen molecule with eight electrons. The OH frequency of hydrogen peroxide with eight electrons is the same as the OH molecule with three electrons which is zero. Placing the number of electrons on an intermedion as a post superscript, the relations described above are summarized with interpolation as:

$$\text{H-OO-H}^{16} = \text{O-H}^{8};\ \text{H-OO-H}^{8} = \text{O-H}^{3} = 0.0\ \therefore \text{H-OO-H}^{14} = \text{O-H}^{6.75} = 3474\ cm^{-1} \qquad (9)$$

$$\text{HO-OH}^{16} = \text{O-O}^{14} = 0.0;\ \text{HO-OH}^{8} = \text{O-O}^{8}\ \therefore \text{HO-OH}^{14} = \text{O-O}^{12.5} = 1347\ cm^{-1} \qquad (10)$$

Thus, when two OH groups are joined to produce hydrogen peroxide, each group looses one-quarter of an electron and the O-O bond gains one-half electron.

To test the calculations in Eqs. (8) and (9), consider the data from.Catalano and Sanborn[24], who obtained spectra of hydrogen peroxide relatively free from hydrogen bonding and other intermolecular complications by suspending small amounts of hydrogen peroxide in a solid nitrogen matrix. The frequencies they reported included 3540, 3417, 1735 and 388 $cm^{-1}$. The last frequency, 388, corresponds to the O-O torsion mode. The combination of O-O stretch and torsion of hydrogen peroxide, that is, 1347 + 388 =1735 is an allowed frequency. The first overtone of this combination band, 2(1735) = 3470, is of the same species as the O-H symmetrical stretch, 3474, and is close enough to produce Fermi Resonance. When two frequencies of the same species are close enough, no band is found at the center but there are two bands symmetrical about the center, 0.5(3540 +3417) = 3478.5 $cm^{-1}$.. The difference between the two resonance bands is inversely proportional to the difference between the two frequencies. **The anomalous frequencies assigned to the symmetrical and unsymmetrical stretch in the spectrum of hydrogen peroxide, are the result of Fermi Resonance.** A more complete discussion of the spectrum of hydrogen peroxide is presented in a previous publication[25].

Returning to catalysis, how many of these intermedions may be formed from a single molecule on a catalyst surface. In the paragraph above which cited the spectra of others from CO and $CO_2$ on nickel and nickel oxide, bands from CO corresponded to CO intermedions with integers of 8, 10 and 11 with fractions about 0.85. Considering the fraction to be 'ε', if there were two species of nickel present, one with an orbital field containing only 'ε' and the other containing one electron and the fraction, '1 + ε' , then depending upon the relative electronegativity of the field and the molecule, there could be no more than four intermedions. The electronegativity of the field would depend not only on the ion but also on the ligands attached to the ion. It is considered that there is no actual transfer of electrons or fractions of electrons between the molecule and the field. Just as if one weighed in alcohol that has a density of 0.8 one would weigh 20% of their weight in air. The process of catalysis is a 'field effect' accomplished by electric fields. This explains why a catalyst is not consumed. The field with only 'ε' is called an 'ic' field and can appear to donate 'ε' or to accept '2 - ε'. The field with '1 + ε' is called an 'ous' field and can appear to donate '1 + ε'or accept '1- ε' electrons. For atoms or molecules with an even number of protons, the 'ic' field corresponds with even integers, etc.

## THE RELATION BETWEEN IONIZATION AND OXIDATION

The preceding paragraph led to the establishment of the following relation between 'ionization' and 'oxidation'. The atom is always the chemist's zero oxidation state whether or not it has low-lying excited electron energy states. Increasing consecutively. each ionization state corresponds to one oxidation state unless that ionization state has low-lying levels. Then it corresponds to a single fraction but to two oxidation states; a lower state as 'ous' and an upper state as 'ic'. For chromium, the atom corresponds to the zero oxidation state and the positive ion to the chemists' 1+ oxidation state. The double positive ion has low-lying states and therefore corresponds to the single fraction, 0.09773, and two oxidation states. The 'ous' state corresponds to the chemists' 2+ and the 'ic' state to the chemists' 3+ oxidation state. The agreement is excellent except for a few cases, such as, nickelous 2+ is an 'ic' oxidation state because the 1+ state was not recognized when the naming occurred. Unfortunately there are negative states which can not be described with the Roman numerals currently in use.

The catalytic site is considered to contain two electric fields, orbitals, which are **not** connected by a bond. However the two electric fields must be close enough to overlap, providing for mass transfer and providing the site for the catalytic event. Both fields must be in the same oxidation state, that is, both 'ic' or both 'ous'. The appropriate ion to catalyze a reaction of A + B ↔ C has a fraction that, when imparted to both reactants, makes the frequencies of the two reactants 'A' and 'B' harmonically equal, that is they differ only by powers of two. **When the reactants meet in the region of overlap, because their frequencies are now equal, resonance occurs in the direction to decrease the Free Energy.. It is during resonance that electrons are shared, paired and exchanged and bonds are made or broken**.

As complete as the foregoing discussion maybe, it only applies to the bonds of molecules. There are three times as many reactions involving an atom in a molecule as there are bond-to-bond reactions. These would be atom in molecule A to bond in molecule B, etc. There is no way at this time to detect the frequency of the electric field at an atom in a molecule or the frequency of the electric field of an individual atom Nevertheless those equations must be derived to complete this explanation of catalysis.

Equation (2) expresses the frequency of the bond as electrons are added to anti-bonding orbitals going from a maximum at eight electrons to zero at twelve electrons..
The frequency in Eq.2, for CO, decreases as electrons are added. This could be considered an "electron accepting capability". There should be an "electron donating capability" that increases as electrons are added One equation[26] is the inverse of the other. The frequencies of the two equations should add to a constant value for any number of electrons To convert from Eq. (2) to the equation at the atom, the factor 'number of electrons' remains the same, the sign of 'k', the curvature is reversed and the asymptote of the frequency factor is: [ $k/(\eta_{min} - B)$], e.g. [268.309/(8 - 12.1182)] =
- 65.1520

$$(\nu_{\text{-co}} + 65.1520)(\eta_{co} - 12.1182) = -268.309\ \text{cm}^{-1} \qquad (11)$$

## PREDICTION OF CATALYSTS

The reaction to be catalyzed must be broken into elementary steps of the form "A + B↔C" in which 'C' must be capable of being formed, i.e. 'C' must have some existence. A pair of hyperbola equations are chosen one from each reactant This defines the reaction mechanism of each elementary reaction. i.e.atom from 'A' with bond from 'B', etc. The '$\eta$' must be replaced by 'I + f' where 'I' is the integers of the intermedion and 'f' is the fraction. Simultaneous solution of pairs of these hyperbolas, one from each reactant, provides the following:

1. The fraction of the number of electrons that makes the frequencies at the sites of reaction harmonically equal identifies the catalyst by a four digit number.
2 The relation of ionization to oxidation allows specification of the oxidation state of the catalyst at the instant of the catalytic event, for catalysts that are positive ions.

3 The integers of the reacting intermedions specify whether the reaction requires a solid catalyst or may occur in a homogeneous system

The AEL tables list only positive ions, in many cases going to ionization states as high as the nucleus, The ions of interest to catalysis, to date, have low-lying electronic states at levels no greater than the sixth. There are metal ions that impart a fraction to surface species that do not show the low-lying electronic levels below the seventh level, such as zinc, gold, sodium and cadmium. It is considered that these must be functioning in catalysis as negative ions. It would be very difficult to generate tables of negative ions of the elements. This would require shooting electrons at the elements with just the energy to land in the lowest vacant orbital. The discussion following the table analyzing the zinc catalyzed oxidation of CO will present some evidence of the presence of negative zinc ions.

## QUANTITATIVE ANALYSES OF CATALYTIC REACTIONS

The equations used in each table are listed. Calculated frequencies are multiplied or divided by powers of two to facilitate comparisons in parentheses. Those that match are bolded, italicized and underlined

**Oxidation of CO catalyzed by Gold (PF = 0.8035)**

$(\nu_{co} - 2269.96)(\eta_{co} - 12.1182) = 268.309\ \text{cm}^{1}$ (2)

$(\nu_{-co} + 65.1520)(\eta_{co} - 12.1182) = -268.309\ \text{cm}^{1}$ (11)

$(\nu_{oo} - 2920.16)(\eta_{oo} - 15.7525) = 5117.58\ \text{cm}^{1}$ (7)

$(\nu_{-oo} + 660.12)(\eta_{oo} - 15.7525) = -5117.58\ \text{cm}^{1}$ (12)

**Table 4: The four CO intermedions and the four $O_2$ intermedions on gold (PF = 0.8035)**

**CO Intermedions on Gold**

| | | | | |
|---|---|---|---|---|
| $\eta_{co}$ | 8.8035 | 9.8035 | 10.8035 | 11.8035 |
| $\nu_{co}$ | 2189.01 | 2154.04 | 2065.88 | 1417.37 |
| | (547.25) | (538.51) | (516.47) | (708.69) |
| | | | | |
| $\nu_{-co}$ | 15.79 | 50.76 | 138.93 | 787.43 |
| | (505.28) | *(**812.16**)* | (555.73) | |

**$O_2$ Intermedions on Gold**

| | | | | |
|---|---|---|---|---|
| $\eta_{oo}$ | 10.8035 | 11.8035 | 12.8035 | 13.8035 |
| $\nu_{oo}$ | 1886.10 | 1624.24 | 1184.80 | 294.41 |
| | (943.05) | (***812.12***) | (592.40) | (588.83) |
| | | | | |
| $\nu_{-oo}$ | 373.943 | 635.798 | 1075.24 | 1965.63 |
| | 747.89 | | 537.62 | 982.81 |

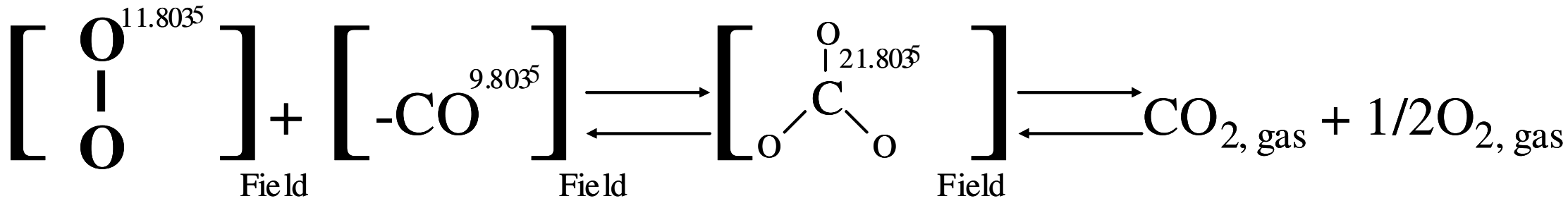

**Figure 2. Schematic of CO oxidation catalyzed by gold compound.**

Oxygen and carbon dioxide are 'even' molecules. The matching intermedions have numbers of electrons with odd integral numbers. Therefore the catalyst is in the 'ous' oxidation state and all even molecules on the catalyst have odd integral numbers of electrons. For 'odd' molecules on a catalyst in the 'ous' oxidation state, the integers are even and so forth.

The two intermedions that match harmonically are in the same column in the two tables. This means that not only are they matched with respect to oxidation state but also with respect to electronegativity. The apparent electron transfer is in the same extent and direction for both of the reactants. These are the requirements for a homogeneous reaction. A 0.001 molar solution of gold cyanide catalyzed the oxidation of carbon monoxide at room temperature. Cyanide ion was chosen as the ligand both to dissolve the gold and because cyanide provides a polymeric solute.

**Oxidation of CO catalyzed by Zinc (PF = 0.3247)**

**Table 5: The four CO intermedions and the four $O_2$ intermedions on zinc (PF = 0.3247)**

**CO Intermedions on Zinc**

| | | | | |
|---|---|---|---|---|
| $\eta_{co}$ | 8.3247 | 9.3247 | 10.3247 | 11.3247 |
| $\nu_{co}$ | 2199.23 | 2173.91 | 2120.36 | 1931.85 |
| | (549.81) | (543.48) | (530.09) | (965.92) |
| | | | | |
| $\nu_{-co}$ | 5.5756 | 30.8938 | 84.4445 | 272.960 |
| | (***713.67***) | (988.60) | (675.56) | (545.92) |

**$O_2$ Intermedions on Zinc**

| | | | | |
|---|---|---|---|---|
| $\eta_{oo}$ | 10.3247 | 11.3247 | 12.3247 | 13.3247 |
| $\nu_{oo}$ | 1977.32 | 1764.39 | 1427.22 | 812.30 |
| | (988.66) | (882.19) | (***713.61***) | |
| | | | | |
| $\nu_{-oo}$ | 282.717 | 495.651 | 832.822 | 1447.74 |
| | (565.43) | (991.30) | | (723.87) |

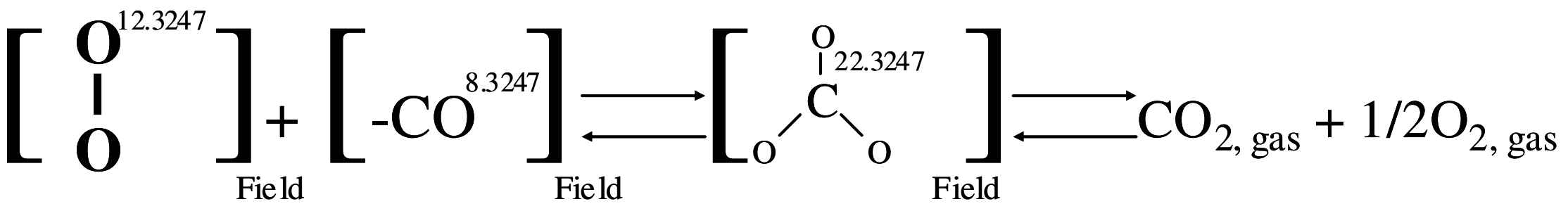

**Figure 3. Schematic of CO oxidation catalyzed by zinc compound**

Taylor et al [14] note that the infrared band at 2174 for CO on zinc oxide slowly diminished and disappeared as a band at 2198 started and grew. The band at 2198 corresponds to the $\nu 3$ band of $CO_2$. Thus zinc catalyzed the oxidation of CO to $CO_2$. The question is why did this reaction take more than half an hour to take place? If, in order for the zinc ion to function as a catalyst, it would require a negative oxidation state, it would have taken quite some time to find two adjacent negative zinc ions on a positive oxide surface. The reacting intermedions are in different columns therefore this reaction can only proceed on a solid catalyst.

## THE CHLORINATION OF CO TO PHOSGENE CATALYZED BY CHROMOUS ION

The destruction of phosgene, $COCl_2$, stored from WWI and WWII, is a challenging project. The Gibbs Free Energy[27] change, $\Delta G^o$ at 298.15K, is -68.72 kJ/mol to form phosgene from CO and $Cl_2$. The destruction of phosgene could be accomplished with catalysis of the oxidation of CO to $CO_2$. Even though $\Delta G^o$ for the decomposition of phosgene is +68.72 kJ/mol, $\Delta G^o$ for the oxidation of CO is -257.23 kJ/mol, so that $\Delta G^o$ for the overall reaction is -188.5 kJ/mol. The problem would be to maintain chromium in the 2+ state in the presence of oxygen.

$$(\nu_{CO} - 2269.96)(\eta_{CO} - 12.1182) = 268.309 \qquad (2)$$

$$(\nu_{-CO} + 65.1520)(\eta_{CO} - 12.1182) = -268.309 \qquad (11)$$

$$(\nu_{Cl_2} - 909.228)(\eta_{Cl2} - 17.2631) = 1148.48 \qquad (13)$$

$$(\nu_{-Cl_2} + 123.984)(\eta_{Cl2} - 17.2631) = -1148.48 \qquad (14)$$

**Table 6: The four CO intermedions and the four $Cl_2$ intermedions on chromium (PF = 0.09773)**

CO Intermedions on Chromium +2 and +3

| | | | | |
|---|---|---|---|---|
| $\eta_{CO}$ | 8.09773 | 9.09773 | 10.09773 | 11.09773 |
| $\nu_{CO}$ | 2203.22 | 2181.13 | 2137.16 | 2007.03 |
| | (550.81) | (545.28) | (534.29) | (501.76) |
| | | | | |
| $\nu_{-CO}$ | 1.5836 | 23.6781 | 67.6427 | 197.775 |
| | (810.81) | (***757.70***) | (541.15) | (791.10) |

**$Cl_2$ Intermedions on Chromium +2 and +3**

| | | | | |
|---|---|---|---|---|
| $\eta_{Cl_2}$ | 12.09773 | 13.09773 | 14.09773 | 15.09773 |
| $\nu_{Cl_2}$ | 686.88 | 633.51 | 546.40 | 378.84 |
| | | | | (***757.68***) |
| | | | | |
| $\nu_{-Cl_2}$ | 98.358 | 151.74 | 362.83 | 406.40 |
| | (786.87) | (606.95) | (725.65) | (812.80) |

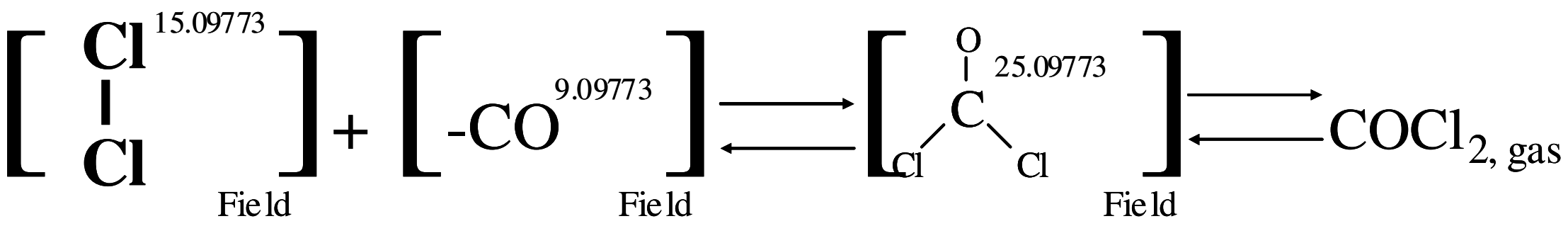

**Figure 4. Schematic of CO chlorination catalyzed by chromium compound.**

## ANALYSES OF CARBONIC ANHYDRASE AND HOMOLOGUES

Carbonic Anhydrase is an enzyme in most of the cells in all animals. Originally it was thought to catalyze the reaction of water with carbon dioxide to form carbonic acid. The carbonic acid was then thought to dissociate forming a proton plus a bicarbonate anion. This reaction converts gaseous $CO_2$ into soluble bicarbonate, decreasing the volume from gas at 24.5 l/mole to liquid at some 0.1 l/mole. This enzyme is estimated to catalyze some 30 million reactions per second. This prevents explosions upon digestion. Northrup[27] through many isotopic and rate studies found that OH was the hydrating reactant. Zinc is the metal in the naturally occurring enzyme. Researchers[28] replaced the zinc with nickel, iron, cobalt and cadmium, all in the +2 oxidation state. The nickel[29] and iron compounds were inactive. The cobalt and cadmium homologues were 10% as active as the zinc.
The equations for the reactions of zinc carbonic anhydrase and homologues

$$(\nu_{OH} - 6090.25)(\eta_{OH} - 0.16649) = -17256.8 \quad (8)$$

$$(\nu_{-OH} + 2202.94)(\eta_{OH} - 0.16649) = 17256.8 \quad (15)$$

$$(\nu_{2CO_2} - 743.34)(\eta_{CO2} - 18.2283) = 169.704 \quad (16)$$

$$(\nu_{2\text{-}CO_2} + 76.16)(\eta_{CO2} - 18.2283) = -169.704 \quad (17)$$

**Table 7. Three of the four OH intermedions and the two $CO_2$ intermedions on zinc (PF = 0.3247)**

**OH Intermedions on Zinc**

| | | | | |
|---|---|---|---|---|
| $\eta_{OH}$ | 5.3247 | 6.3247 | 7.3247 | |
| $\nu_{OH}$ | 2744.74 | 3288.00 | 3679.48 | |
| | (686.11) | (822.00) | (919.87) | |
| | | | | |
| $\nu_{-OH}$ | 1142.57 | 599.31 | 207.83 | |
| | (571.28) | | (***<u>831.33</u>***) | |

**$\nu_2CO_2$ Intermedions on Zinc****

| | | | | |
|---|---|---|---|---|
| $\eta_{CO2}$ | | | 16.3247 | 17.3247 |
| $\nu_2CO_2$ | | | 654.22 | 555.53 |
| | | | | |
| | | | | |
| $\nu_2$-$CO_2$ | | | 12.989 | 111.65 |
| | | | ***<u>(831.30)</u>*** | |

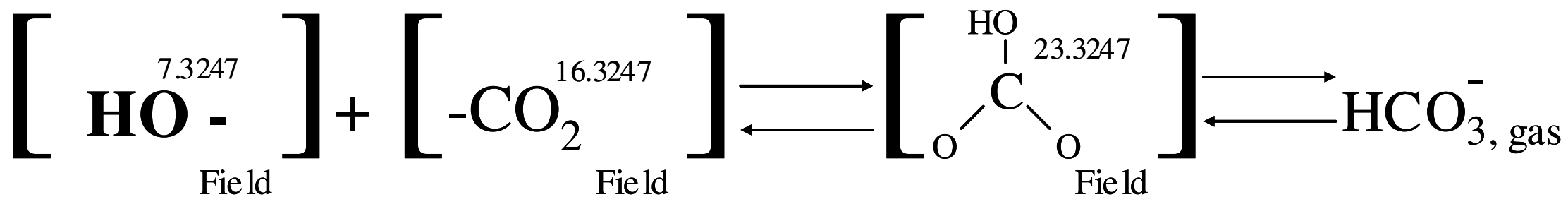

**Figure 10. Schematic of hydration of carbon dioxide catalyzed by zinc compound.**

** The $\nu_2CO_2$ intermedions in the tables for carbonic anhydrase reactions are placed to the right since both correspond to apparent electron donation by the catalyst.
The data in Table 7 defines the reaction mechanism at the instant of the catalytic event as the attack of the carbon atom of the 15.3247 $CO_2$ intermedion onto the oxygen of the 7.3247 OH intermedion forming $HCO_3$ in a single step. This analysis clearly supports Northrup's conclusion that the hydrating agent is OH instead of $H_2O$.

**Table 8. Three of the four OH intermedions and the two $CO_2$ intermedions on cobalt (PF = 0.6221)**

OH Intermedions on Cobalt +1 and +2

| | | | | |
|---|---|---|---|---|
| $\eta_{OH}$ | 5.6221 | 6.6221 | 7.6221 | |
| $\nu_{OH}$ | 2927.12 | 3417.10 | 3775.64 | |
| | (731.78) | (854.28) | (***<u>943.91</u>***) | |
| | | | | |
| $\nu_{-OH}$ | 960.19 | 470.21 | 111.67 | |
| | | (940.42) | (893.36) | |

**$\nu_{2CO_2}$ Intermedions on Cobalt**

| | | | | |
|---|---|---|---|---|
| $\eta_{CO2}$ | | | 16.6221 | 17.6221 |
| $\nu_{2CO_2}$ | | | 637.68 | 463.39 |
| | | | | (926.79) |
| | | | | |
| $\nu_{2\text{-}CO_2}$ | | | 29.50 | 203.79 |
| | | | (***<u>943.86</u>***) | (815.16) |

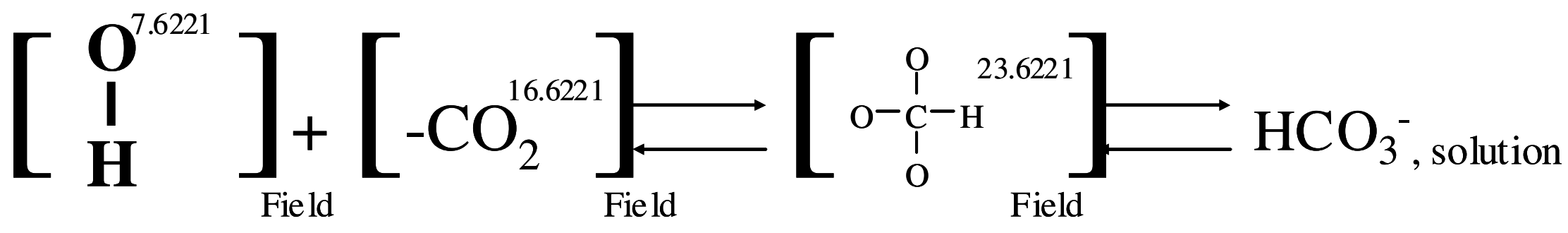

**Figure 11. Schematic of hydration of carbon dioxide catalyzed by cobalt compound.**

**Table 9. Three of four OH intermedions and the two $CO_2$ intermedions on cadmium (PF = 0.3546)**

## OH Intermedions on Cadmium

| | | | | |
|---|---|---|---|---|
| $\eta_{OH}$ | 5.3546 | 6.3546 | 7.3546 | |
| $\nu_{OH}$ | 2764.03 | 3301.55 | 3689.51 | |
| | (691.01) | (825.39.28) | (***<u>922.38</u>***) | |
| | | | | |
| $\nu_{-OH}$ | 1123.28 | 585.76 | 197.80 | |
| | (561.64) | | (791.20) | |

## $\nu_2CO_2$ Intermedions on Cadmium

| | | | | |
|---|---|---|---|---|
| $\eta_{CO2}$ | | | 16.3546 | 17.3546 |
| $\nu_2CO_2$ | | | 652.76 | 549.10 |
| | | | | |
| | | | | |
| $\nu_2\text{-}CO_2$ | | | 14.412 | 118.08 |
| | | | ***<u>(922.34)</u>*** | (944.64) |

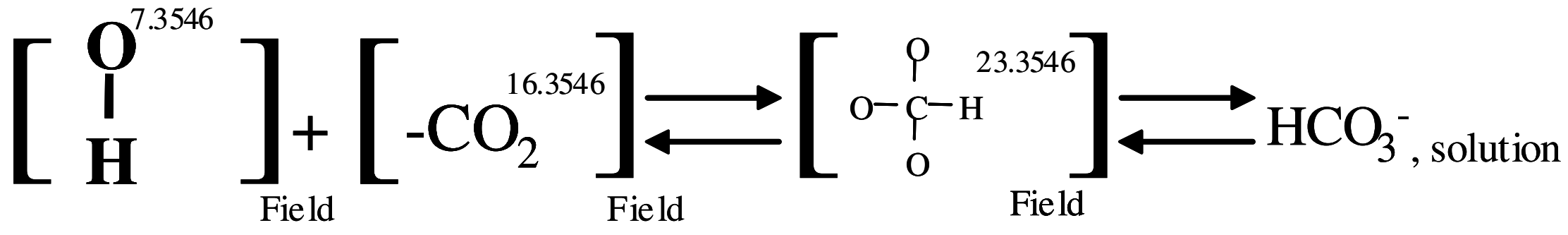

**Figure 12. Schematic of hydration of carbon dioxide catalyzed by cadmium.**

**ONE OF THE MANY WAYS PROTEINS ASSIST CATALYSTS**

The catalytic reaction forming bicarbonate from OH and $CO_2$ provides an excellent opportunity to discuss the methods by which proteins assist enzymes. At the rate of thirty million reactions per second[28] for zinc carbonic anhydrase, this process must represent something close to the maximum reaction rate. A rate this rapid would not allow much if any time at the catalytic site for orienting the two reactants so that the oxygen of the hydroxyl is adjacent to the carbon of the carbon dioxide. Thus the function of the proteins is to orient and present pairs of the reactants to the point of overlap of the two electric fields of the catalyst. The orientation and presentation must be accomplished by regions of appropriate polarity along proteins leading to the catalytic site. This process must represent the epitome of nanotechnology. The slightest error would produce a molecular "jam" of great proportions.

For both cobalt and cadmium enzymes the reaction at the instant of the catalytic event is attack of the carbon of the $CO_2$ **into the bond** of OH forming an unstable intermediate, a carbon with three oxygen atoms and a hydrogen atom each individually attached. Immediately the mobile proton would attach to one of the oxygen atoms forming bicarbonate. It is of interest that the attack of the carbon of $CO_2$ is shifted ever so slightly from the oxygen of the hydroxyl to the O-to-H bond of the hydroxyl. This shift may be less than a femtometer and not recognized by any other method of analysis. The cobalt and cadmium homologues must be able to use the same set of proteins but with significantly decreased efficiency.. Apparently the slight shift of the attack of the carbon is still accommodated by the proteins but with a 90% decrease in efficiency.

## NUMERICAL ANALYSIS OF ELIMINATION OF NITRIC OXIDE

The final catalytic reaction to be considered is the reaction to decrease nitric oxide in automobile exhaust. The following equations of intermedions are used

$$(\nu_{\text{-co}} + 65.1521)(\eta_{\text{co}} - 12.1182) = -268.309\text{cm}^{-1} \quad (11)$$

$$(\nu\text{-n}_2 + 64.6528)(\eta_{\text{n}_2} - 12.1081) = -265.598\text{cm}^{-1} \quad (18)$$

$$(\nu\text{-no} + 142.368)(\eta_{\text{no}} - 13.3459) = -761.078\text{cm}^{-1} \quad (19)$$

$$(\nu_2\text{-nco} + 104.300)(\eta_{\text{nco}} - 17.4250) = -357.231\text{cm}^{-1} \quad (20)$$

In addition to palladium and platinum to catalyze the oxidation of CO to $CO_2$, the automotive catalytic converter contains a trace of rhodium to eliminate nitric oxide (NO) from the exhaust. Quite unexpectedly, nitrogen ($N_2$), which is frequently used as an inert gas, is a major reactant in the following steps to destroy NO.

1. $N_{2cat} + CO_{cat} \leftrightarrow NNCO_{cat}$ catalyzed by rhodium, fraction 0.7965 (21)
2. $NNCO_{cat} \leftrightarrow N_{gas} + NCO_{gas}$ dissociates on leaving catalyst field (22)
3. $NO_{gas} \leftrightarrow NO_{gas}^{+} + e^{-}$ looses an electron at elevated temperature (23)
4. $NO_{gas}^{+} + \nu_2\text{- } NCO_{gas} \leftrightarrow ONNCO_{gas}^{+}$ react upon collision without catalyst (24)
5. $ONNCO_{gas}^{+} + e^{-} \leftrightarrow N_{2,gas} + CO_{2,gas}$ addition of $e^{-}$ dissociates ion to products (25)

Overall this is a redox reaction, CO is oxidized and NO is reduced.

The integers for both $N_2$ and CO are 8,9,10 and 11.

**Table 10. Four $N_2$ intermedions and four CO intermedions on rhodium (PF = 0.7965)**

**$N_2$ Intermedions on Rhodium,**

| | | | | |
|---|---|---|---|---|
| $\eta_{N_2}$ | 8.7965 | 9.7965 | 10.7965 | 11.7965 |
| $\nu_{N_2}$ | 2377.45 | 2342.75 | 2255.15 | 1605.28 |
| | (594.362) | (585.688) | (563.788) | (802.641) |
| | | | | |
| $\nu_{-N_2}$ | 15.5495 | 50.2451 | **137.846** | 787.716 |
| | (995.169) | (803.922) | | |

**CO Intermedions on Rhodium,**

| | | | | |
|---|---|---|---|---|
| $\eta_{co}$ | 8.7965 | 9.7965 | 10.7965 | 11.7965 |
| $\nu_{co}$ | 2189.19 | 2154.39 | 2066.96 | 1435.93 |
| | (547.296) | (538.599) | (516.739) | (717.962) |
| | | | | |
| $\nu_{-co}$ | 15.6225 | 50.4136 | **137.851** | 768.883 |
| | (999.840) | (806.618) | | |

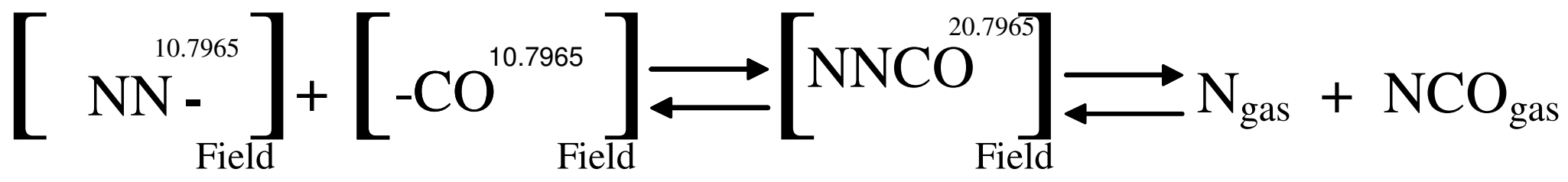

**Figure 3 Schematic of reaction of $N_2$ and CO catalyzed by rhodium.**

The match of $\nu_{-N_2}$ with $\nu_{-co}$ at both atoms indicates the formation of NNCO in the catalyst field. The NNCO molecule dissociates upon leaving the catalyst field into $N_{gas}$ + $NCO_{gas}$ in the second reaction. The loss of an electron in the third reaction would occur readily at the elevated temperature of the auto exhaust. Reaction 4 is a non-catalytic combination reaction at the nitrogen atoms of the neutral molecule, $\nu_2$ NCO and the positive ion, $NO^+$. The frequencies at the nitrogen of each reactant are:

| $^+NO^{10}$ | $\nu_2 NCO^{15}$ |
|---|---|
| 85.898 | 42.949 |
| | (85.898) |

Of the five reactions, only reactions 1 and 4 are expressed numerically. The preceding five reactions explain why the oxygen sensor in an automobile must not be set to oxidize all of the CO to $CO_2$ but must allow some to remain so that CO is oxidized to $CO_2$ while NO is reduced to $N_2$. The $N_2$ and CO intermedions that react in Table 10 not only correspond to the 'ic' oxidation state but also to the same electronegativity in apparent electron transfer, that is they are both in the same vertical column. Thus this reaction is predicted to occur not only in a heterogeneous system such as a catalytic converter but also in a homogeneous system. Since rhodium does not show low-lying levels in the Atomic Energy Level tables, it is assumed to function as a catalyst as a negative ion. Therefore, the cation or ligand to be used should be an electron-donating entity such as chloride ion. The temperature of the reaction should be high enough to cause loss of an electron from nitric oxide and the medium should be an insulator to stabilize the electron. The following prediction may be made for this reaction using a solid catalyst. Placing a very small amount of rhodium chloride with equal partial pressures of CO and NO in a container will result in no reaction. Adding a small amount of nitrogen will start a reaction that will cause a decrease in pressure to completion. The final pressure will be 75% of the initial pressure and the container will contain only $N_2$ and $CO_2$.at partial pressures of 1/3 and 2/3 respectively. If the rhodium chloride is a solid, there should be only two questions, namely, is the temperature high enough to cause NO to readily lose an electron and is the rhodium in the proper oxidation state, probably +3. If rhodium functioned as a catalyst as a positive ion and the data were available in the AEL tables, the oxidation state could be specified exactly.

## LIST OF FRACTIONS AND CORRESPONDING IONS

a-derived from experiment, data not available in Atomic Energy Levels tables
b-confirmed experimentally and calculated from Atomic Energy Levels tables
IIICr 0.09773b; IIFe 0.2587b;Cu 0.3241a; Zn 0.3247a; Cd 0.3546a;Ca 0.4964a; Ag 0.5670a; IICo 0.6221b; IIIRu 0.7416b; Rh 0.7965a; Au 0.8035a; Pd 0.8120a; IIINi 0.8595b; Pt 0.8662a; IIRu 0.9420b:

## CONCLUSION

Figure 1 relating data from the Atomic Energy Level tables provides definitive proof of the origin of the fractions assigned to the catalytic metals. All fractions observed since have fit upon this single straight line. The discovery that electric fields with fractional charges are the catalysts answers the 160-year old quest for an understanding of the mechanism of this phenomenon. Approaching the problems of the numerical expression of catalytic and non-catalytic chemical reactions in terms of electric fields adds a non-exclusive perspective that will complement other current paradigms. Recognition of a third type of interaction of a gas with a solid surface has led to new interpretations of chemical and catalytic reactions, and the discovery of 'intermedions' the intermediates of catalysis. A right hyperbola of the form

$(\nu - A)(\eta - B) = k$, wherein 'ν' is frequency, 'η' is number of electrons, 'A and B' are asymptotes, and 'k' curvature specific for each atom and bond in each molecule converts spectral bands into numbers of electrons. This conversion greatly increases the information to be obtained from IR spectroscopy.

The unsolved problem of the anomalous assignments of the symmetrical and unsymmetrical OH stretches in hydrogen peroxide are shown to be due to Fermi Resonance. The resonance is between the single value of both the symmetrical and the unsymmetrical OH stretches with the first overtone of the combined OO stretch and torsion modes.

**ACKNOWLEDGEMENTS**
I wish to acknowledge the encouragement and constructive information provided by, Professors Luigi Messineo, Jearl Walker, David Anderson and my Thesis Advisor, Ralph Petrucci, at Western Reserve University and credit James Thomas, Jr. with the suggestion that the activity across a cell wall may also involve electric fields.

**Appendix**
There were a few examples that may have been thought to be intermedions that did not agree either in frequency or number of electrons with others. These "outliers" in the case of the experiments of other investigators may have been due to reports of frequencies before evacuation rather than the frequencies removed by evacuation. These conditions may not have been important to others. Another reason for "outliers" may have been due to the fact that all chemical bonds are not of the same strength. The nickel to carbon bond in nickel tetracarbonyl breaks at 300 degrees centigrade while the carbon to oxygen bond does not. Since every attempt was made to maximize the variety of intermedions in the first experiments by using Cab-o-sil, an aerogel of surface area of some 400 square meters per gram there may have been one or two cases of a carbon to silica bond at a corner or edge of a crystal of $SiO_2$ that was weak enough to break at ambient temperature. The frequency of the carbon to oxygen bond of such a carbonyl or the number of electrons calculated by Eq..2 would be unlikely to agree with the intermedions.